# On the magnetoresistance anisotropy of a 2-dimensional electron gas with large half-integer filling factors

Herbert Kroemer[*]

Department of Electrical and Computer Engineering,
and QUEST, Center for Quantized Electronic Structures,
University of California, Santa Barbara CA 93106

**Abstract**

The unsymmetric confinement of the electrons in a typical 2-dimensional electron gas (2DEG) breaks the full cubic symmetry of the in-plane transport properties. A *rigorous* invariance of the transport under rotation by π/2 about the cubic axis perpendicular to the plane of the gas no longer holds, even for structurally perfect samples, and asymmetries under rotation by π/2 are allowed. It is proposed that this symmetry breaking plays a role in the recent observation, by Little et al., of the pronounced anisotropy of the magnetoresistance of a 2DEG at very low temperatures, at large half-integer filling factors of the Landau levels in such a system.

## I. INTRODUCTION

In a recent paper, Lilly et al. [1] have presented strong evidence for an unexpected dramatic anisotropy in the low-temperature (≤150mK) magnetoresistance of a high-quality two-dimensional electron gas (2DEG), when the strength of the magnetic field is such that the Landau level filling factor is a half-integer of the form $v = \mu + 1/2$, where $\mu$ is an integer ≥ 4. More specifically, under those conditions, the resistances encountered by currents flowing in the crystallographic $[110]$ and $[1\bar{1}0]$ directions differ drastically: contrary to expectations, the resistance is not invariant under a rotation by π/2 about the cubic $[001]$ axis. There are no plateaus in the Hall resistance associated with the magnetoresistance anomaly.

With respect to potential explanations of the phenomenon, the authors point out that it is not at all clear what breaks the in-plane symmetry of the system. The authors apparently favor a mechanism based on the spontaneous formation of charge density waves (CDWs) leading to in-plane stripes of alternating charge density. Such a phenomenon could clearly lead to a huge anisotropy. However, this explanation still poses the question of what determines the preferred orientation of the charge density stripes—which would have to be coherent over the macroscopic size of the samples. A possible source might be built-in structural imperfections in the samples, especially steps at the GaAs-(Al,Ga)As hetero-interface due a slight (and almost unavoidable) deviation of the interface from a perfect (100) plane.

---

[*] kroemer@ece.ucsb.edu

The purpose of the present note is to point out that the symmetry arguments for a full cubic isotropy in homogeneous *bulk* GaAs break down inside a 2DEG of the kind studied by Lilly et al., and that there are no rigorous symmetry objections against a loss of full cubic symmetry of the in-plane transport properties. This symmetry breaking may be what causes the preference for a certain orientation of the postulated CDW stripes. Less likely, it might also cause a transport anisotropy all by itself.

## II. SYMMETRY BREAKING BY ASYMMETRIC CONFINEMENT

Our point of departure is that the bulk crystal structure of GaAs is not invariant under *pure* rotations by $\pi/2$ about the cubic axes, but only under rotation by $\pi$. What causes the full cubic invariance of the bulk *transport properties* is the invariance of the crystal structure under a *rotary reflection* $S_4$ that combines a rotation by $\pi/2$ about the cubic axes with a simultaneous reflection at a plane perpendicular to that rotation axis [2]. Alternatively, $S_4$ may be viewed as a combination of a rotation by $-\pi/2$ and the inversion. None of these operations are by themselves symmetry elements of the bulk structure; only their combinations are. But it is well-known that the electron energy function $E(\mathbf{k})$ in $\mathbf{k}$-space always has inversion symmetry, due to time-reversal invariance, even if the crystal structure does not have inversion symmetry. Hence, if $S_4$ is present in *real* space, the rotations by $\pm\pi/2$ also become symmetry invariances of $\mathbf{k}$-space by themselves, even if they are not invariances in real space. This leads to *full* cubic symmetry of $\mathbf{k}$-space, which in turn implies full cubic symmetry of the *bulk* transport properties.

This bulk symmetry argument breaks down inside a 2DEG of the kind studied by Lilly et al. In such a 2DEG, the electrons are confined to an unsymmetric quantum well, with an abrupt heterobarrier on one side, and an electric field on the order $2\text{-}4\times10^4$ V/cm on the other. Neither the abrupt wall nor the electric field are invariant under reflection at any plane parallel to the plane of the 2DEG itself. But this means that the 2DEG is not invariant under $S_4$. In this case the conclusion can no longer be drawn that the in-plane properties of the 2DEG should be rigorously invariant under rotation by π/2.

To see how an anisotropy under π/2 rotation might arise, consider the geometry of the bonding and anti-bonding orbitals connecting a Ga atom to its four As nearest neighbors. If we pick our cubic axis system in such a way that the coordinate origin coincides with the Ga atom, and that the [111] direction is the direction to one of the four nearest-neighbor As atoms, we can distinguish two different sets of bonding or anti-bonding orbitals: One set, here called {+}, consists of the orbitals oriented along $[111]$ and $[\bar{1}\bar{1}1]$, with a "+1" for the third Miller index. The other set, {–}, consists of the orbitals oriented along $[1\bar{1}\bar{1}]$ and $[\bar{1}1\bar{1}]$.

In the presence of an unsymmetric confinement, both sets of orbitals will become polarized in a way that destroys their symmetry equivalence under $S_4$. For example, a force acting on the electrons in the +z direction (as is present on the GaAs side of the hetero-interface) will presumably polarize the bonding orbitals in {+} towards the As atoms, adding to the polarization already present from the Ga-As asymmetry, and increasing the net polarization of these orbitals. For the bonding orbitals in the set {–} the opposite is true. The anti-bonding orbitals, which are more relevant here, because they make up the conduction band states, behave oppositely to the bonding orbitals.

As was pointed out earlier, the CDW hypothesis poses the question of what determines the preferred orientation of the charge density stripes. The broken in-plane cubic symmetry of the 2DEG might very well be the sought-after mechanism, providing the missing link in the CDW theory.

Little et al. report that a sharp peak in the resistivity $\rho_{xx}$ occurs for current flow in the $[1\bar{1}0]$ direction, implying CDW stripes parallel to $[110]$. This preference must somehow be related to the different orientations of the two inequivalent sets of orbitals. Consider the orbitals of the set {+}, together with their translational equivalents in other crystallographic cells. They evidently connect to form a set of continuous zigzag strings along the $[110]$ direction, the direction of the CDW stripes, but with little overlap between parallel strings. For the set {–}, the orbitals connect in the $[1\bar{1}0]$ direction, perpendicular to the CDW stripes. Presumably, it is the inequivalence of the two sets of orbitals that causes CDW stripes to form in the $[110]$ direction, the direction of the strings of {+} orbitals. Just exactly why the preference has this particular direction, rather than the orthogonal one, is a quantitative question that goes beyond the limited scope of the present short note, and which has not been attempted.

## III. DO WE NEED INTERACTION EFFECTS?

Given the broken symmetry, the question naturally arises whether it cannot explain the observations by itself, as an elementary "ordinary" form of conduction anisotropy, without the amplification via the nucleation of CDWs with a preferred orientation.

The breakdown of the symmetry equivalence of the two sets {+} and {–} of localized orbitals implies a similar breakdown of the symmetry equivalence of pairs of de-localized Bloch waves with mutually orthogonal in-plane **k**-vectors. In particular, the Bloch waves with **k**-vectors in the $\pm[110]$ directions will no longer be symmetry-equivalent to those with **k** parallel to $\pm[1\bar{1}0]$. In terms of the energy function $E(\mathbf{k})$, the original near-circular energy contours with their rotational invariance by $\pi/2$ will acquire an elliptical distortion, which destroys this invariance and which implies, of course, an corresponding reduced symmetry in the transport properties.

The trouble with this seemingly simple explanation is that the observed anisotropies can be huge; resistance ratios close to 100:1 have been reported [1]. This would require a similarly large elliptical distortion of the in-plane energy contours, and I see no way how the comparatively weak symmetry-breaking confinement force could lead to such a huge effect. Furthermore, a simple anisotropy of the energy contours would not lead to a magnetic field dependence such that the anisotropy, after first increasing with decreasing filling factor, abruptly disappears once the filling factor drops below $v = 4$. Evidently, a more complicated process is involved.

This leaves the CDW instabilities proposed by Lilly et al. as the prime candidate for the observed anisotropies—provided a mechanism for inducing a preferred direction for the CDW stripes can be identified. The symmetry breaking induced by the unsymmetric confinement of the 2DEG would provide such a mechanism.